\documentstyle[aas2pp4]{article}
\def\un#1#2{\hbox{\,#1$^{#2}$}}
\def\sub#1{_{\rm{}#1}}
\def\Xray{\hbox{X-ray}}
\def\CenX{\hbox{Cen X-}}
\def\oao{\hbox{OAO~1657$-$415}}
\def\fouru{\hbox{4U~1626$-$67}}
\def\GX{\hbox{GX~1+4}}
\def\pHzpersec{\un{pHz}{}\un{s}{-1}}
\let\simgt\gtrsim\let\simlt\lesssim

\let\internalcite\cite
\def\cite{\def\citename##1{##1}\internalcite}
\def\ctyr{\def\citename##1{}\internalcite}

\makeatletter
\def\thebibliography{\subsection*{REFERENCES}
\list{}{\labelwidth3em\leftmargin\labelwidth\labelsep\z@\parsep\z@
\itemsep\z@\itemindent-3em\usecounter{enumi}}
\def\refpar{\relax}\def\newblock{\hskip .11em plus .33em minus .07em}
\sloppy\clubpenalty4000\widowpenalty4000\sfcode`\.=1000\relax}
\makeatother

\begin{document}

\title{Warped Disks as a Possible Origin of Torque Reversals in\protect\\
Accretion-Powered Pulsars}
\righthead{Warped Disks and Torque Reversals in Accreting Pulsars}
\author{M. H. van Kerkwijk\altaffilmark{1}, 
        Deepto Chakrabarty\altaffilmark{2}, 
        J. E. Pringle\altaffilmark{1,3},
        and
        R. A. M. J. Wijers\altaffilmark{1}}
\altaffiltext{1}{Institute of Astronomy, Madingley Road, Cambridge
             CB3~0HA, UK; mhvk, jep, ramjw@ast.cam.ac.uk}
\altaffiltext{2}{Center for Space Research, Massachusetts Institute of
             Technology, Cambridge, MA 02139, USA;
             deepto@space.mit.edu}
\altaffiltext{3}{Isaac Newton Institute for Mathematical Sciences,
                 20~Clarkson Road, Cambridge CB3 0EH, UK}

\begin{abstract}
Enigmatic transitions between spin-up and spin-down have been observed
in several \Xray\ pulsars accreting matter via an accretion disk.  In
these transitions, the torque changes sign but remains at nearly the
same magnitude.  It has been noted previously that alternating
prograde and retrograde disk flows would explain many features of the
torque reversals, although it has been unclear how a stable retrograde
disk could be formed.  We suggest that the reversals may be related to
the disk at times being warped to such an extent that the inner region
becomes tilted by more than 90 degrees.  This region would thus become
retrograde, leading to a negative torque.  Accretion disk models can
show such behavior, if account is taken of a warping instability due
to irradiation.  The resulting `flip-overs' of the inner parts of the
disk can reproduce most characteristics of the observations, although
it remains unclear what sets the timescale on which the phenomenon
occurs.  If this model were correct, it would have a number of
ramifications, for instance that in the spin-down state the \Xray\
source would mostly be observed through the accretion disk.
\end{abstract}


\section{Introduction}\label{sec:intro}

Long-term, continuous monitoring by the BATSE all-sky monitor on the
{\it{}Compton Gamma Ray Observatory} has led to a qualitative change
in our picture of the spin-frequency behavior of accreting \Xray\
pulsars (Bildsten et al.\ 1997, hereafter \cite{bild&a:97}).  Of the
four well-measured persistent sources thought to accrete by way of an
accretion disk, all display sudden transitions between episodes of
steady spin-up and spin-down.  The spin change rate (i.e., the
absolute value of the pulse frequency derivative) is nearly equal for
spin-up and spin-down, and it is comparable to the torque expected if
all the angular momentum of the accreting gas is deposited at the
magnetospheric boundary and transferred to the neutron star.  In at
least some of these systems, however, Roche-lobe overflow is thought
to occur, and hence the matter inserted into the accretion disk has
only one sense of angular momentum; symmetric torque reversals would
thus not be expected.

Previous accretion torque models, in which the net torque could be
positive or negative depending on the mass accretion rate $\dot M$
(e.g., \cite{ghosl:79}), cannot easily reproduce the observations (see
\cite{nels&a:97}).  We only list the main arguments here: (i) one
needs step-wise changes in $\dot{M}$ to produce distinct spin-up and
spin-down states -- this seems unlikely; (ii) one would expect changes
in $\dot{M}$ to be reflected in the \Xray\ luminosity $L\sub{X}$ --
there are variations in $L\sub{X}$, but these appear uncorrelated with
spin-up/down state; and (iii) one expects at all times a positive
correlation between torque and luminosity -- for \GX\ an
anti-correlation is observed during its spin-down state
(\cite{chak&a:97b}).

Suggestions about a possible cause for the torque reversals have been
made by Yi, Wheeler \& Vishniac (\ctyr{yiwv:97}) and Nelson et al.\
(\ctyr{nels&a:97}).  Yi et al.\ suggested that the reversals were due
to small changes in $\dot{M}$ around a critical value at which the
system changes from a primarily Keplerian flow to a substantially
sub-Keplerian, radially advective flow.  This addresses points (i) and
(ii), but not point (iii).

Nelson et al.\ (\ctyr{nels&a:97}) explored the possibility of having
systems in which nothing changes except the sense of rotation of the
disks.  They found that this would explain the observations very well.
They quoted a suggestion by Makishima et al.\ (\ctyr{maki&a:88}), that
in \GX\ one might have accretion from a wind and thus form a
retrograde disk more easily.  As Nelson et al.\ noted, however, it is
very hard to imagine how a stable retrograde disk could form,
especially in the ultracompact binary \fouru, for which all
indications are that mass transfer is by Roche lobe overflow from a
very low-mass companion.

Here, we suggest a modification of the Nelson et al.\ picture, viz.,
that it is only the inner part of the disk that is changing its sense
of rotation, as a consequence of very strong warping of the disk.
Accretion disks are unstable to warping if lit strongly by a central
radiation source (\cite{prin:96}).  In numerical simulations which
include such irradiation, the disk can sometimes become more and more
warped, until the inner part has become inclined by well over 90
degrees (\cite{prin:97}; see Fig. 6 of that paper for illustrations of
a disk with its inner parts flipped over).

In Section~\ref{sec:obs}, we give an updated summary of the
observations.  In Section~\ref{sec:warp}, we briefly discuss the
warping process and describe simulations done specifically for \Xray\
binaries.  We proceed to make a qualitative comparison with the
observations, and to discuss ramifications.  We summarize our
conclusions in Section~\ref{sec:conc}.

\section{Summary of the Observations}\label{sec:obs}

BATSE has monitored five persistent, disk-fed accreting \Xray\ pulsars
for nearly 7 years (\cite{bild&a:97}).  Here, we summarize the
observations for four of these. We exclude Her~X-1, since it cannot be
monitored continuously, but only during the high states of its 35-d
precession cycle.

{\em 4U 1626$-$67} is a 7.66\,s pulsar in an ultracompact 42\,min
binary.  It has $L\sub{X}\simeq10^{37}\un{erg}{}\un{s}{-1}$
(\cite{chak:98}).  \Xray\ timing measurements suggest we are viewing
the binary nearly face-on.  The $<\!0.1\,M_\odot$ mass donor probably
fills its Roche lobe.  The torque history is reviewed in detail by
Chakrabarty et al.\ (\ctyr{chak&a:97a}). The pulsar underwent steady
spin-up at a mean rate of $0.85\pHzpersec$ during 1977--1990, and
steady spin-down at a mean rate of $-0.72\pHzpersec$ since that time.
The 1990 reversal itself was not observed, but evidently occurred on a
time scale $\lesssim\!1$ month.  Both the spin-up and the spin-down
had a significant quadratic trend.

The \Xray\ luminosity did not change abruptly at the torque reversal,
although there is evidence for a gradual, factor of 2 decay over an
interval of 15 years.  However, the long-term (years) 1--20\,keV X-ray
spectral shape did change, from $dN/dE\propto{}E^{-1.5}$ during
spin-up to $dN/dE\propto{}E^{-0.4}$ during spin-down
(\cite{vaugk:97}), while the $>\!20\un{keV}{}$ spectrum remained
unchanged.  During spin-up, the pulsar showed strong flares on
$\sim\!1000\un{s}{}$ time scales, but these flares are absent or much
less frequent during spin-down (Chakrabarty et al. 1998, in
preparation).

Quasi-periodic oscillations (QPO) were observed both during spin-up
(0.041\,Hz; \cite{shin&a:90}) and spin-down (0.048\,Hz;
\cite{ange&a:95}).  The detailed sideband structure of the X-ray
pulsations and the QPO during spin-down suggests that there is X-ray
reprocessing or scattering in material on a prograde orbit, likely at
small radius (Kommers, Chakrabarty, \& Lewin \ctyr{kommcl:98}).

{\em GX 1+4} is a 2\,min pulsar in a wide
($P\sub{orb}\gtrsim100\un{d}{}$) binary with an M giant, and its
accretion disk is probably fed by the M giant's wind
(\cite{chakr:97}).  Most likely, 
$L\sub{X}\simeq10^{37}\un{erg}{}\un{s}{-1}$, although a larger
distance corresponding to $L\sub{X}\simeq10^{38}\un{erg}{}\un{s}{-1}$
cannot be not ruled out.  The torque and flux histories are reviewed
in detail by Chakrabarty et al.\ (\ctyr{chak&a:97b}).  The source was
bright throughout the 1970s and spinning up rapidly at a mean rate of
$+6.0\pHzpersec$.  {\it{}EXOSAT} observations in 1983 and 1984 failed
to detect it.  When detected again at a low luminosity in 1987, it was
spinning down at a mean rate of $-3.7\pHzpersec$ (1989--1991).  BATSE
has observed continued spin-down since 1991, apart from a short
spin-up episode and an episode in which the source was not detectable.
The torque reversals observed by BATSE were resolved in 5-day-averaged
spin measurements.

The long-term mean \Xray\ luminosity was brighter during the 1970s
spin-up interval (and also during the BATSE spin-up event) than during
the spin-down intervals in the 1980s and 1990s.  However, during the
extended spin-down interval observed by BATSE, an anti-correlation
between torque and \Xray\ luminosity was found: increased spin-down
for increased \Xray\ flux.  The luminosity during spin-down
occasionally flared to values similar to those measured during spin-up
in the 1970s.  The X-ray spectrum did not change drastically between
the 1970s and the late 1980s (\cite{saka&a:90}), although considerable
variability in the hydrogen column density has been observed.

{\em Cen X-3} is a 4.8\,s pulsar in a 2.1\,d binary with an OB
supergiant.  It has $L\sub{X}\simeq10^{37}\un{erg}{}\un{s}{-1}$.
\CenX3 has shown a long-term spin-up trend of $1.6\pHzpersec$ since
1972, but BATSE observations resolve this trend into alternating
episodes of spin-up and spin-down on 10--100 day time scales
(\cite{fingwf:94}).  The torque distribution is bimodal, with
preferred spin-up/spin-down rates of $+7$ and $-3\pHzpersec$
(\cite{bild&a:97}).  The torque reversals are unresolved in the BATSE
measurements and must occur in $<\!10$\,d.  Over a factor of 6 range
in luminosity, {\em no} correlation between torque and luminosity is
found, and the mean flux is comparable in the two torque states
(\cite{vaug&a:98}).  The spectral shape, pulse profile, and pulsed
fraction are correlated with luminosity.

{\em OAO 1657$-$415} is a 38\,s pulsar in a 10.4\,d binary with an
unidentified companion which is inferred to be an OB supergiant.  It
has $L\sub{X}\simeq10^{37}\un{erg}{}\un{s}{-1}$ (\cite{chak&a:93}).
There is no direct evidence for an accretion disk, but the timing
properties of the pulsar strongly suggest disk accretion.  Like
\CenX3, the pulsar has shown a long-term spin-up trend since 1979 of
$\sim\!1\pHzpersec$, but BATSE observations show alternating episodes
of spin-up and spin-down on 10--100\,d time scales, with typical rates
of +7 and $-2\pHzpersec$ (\cite{bild&a:97}).  The torque reversals are
unresolved in the BATSE measurements.  No correlation between torque
and luminosity is observed, although this has only been tested over a
small range (factor of 2) in luminosity (\cite{koh:98}).

\section{Warped disks}\label{sec:warp}

Wijers and Pringle (1997; hereafter WP97) have investigated in detail
the behavior of irradiated accretion disks in binary stars subject to
internal viscous forces, radiation reaction due to illumination from
the central source, and the tide from the mass donor.  The main aim
was to see whether warped disks could form, and whether their
precession could lead to the superorbital periodicities observed in
many \Xray\ binaries.  Here, we summarize the results relevant to the
torque reversals, and compare them to the observations.

Whether the disk becomes unstable to warping depends on the parameter
$F_*$, given by (Eq.~18, WP97)
\[
  F_*=0.127\,\,\eta\,
      \frac{\epsilon}{0.1}\,
      \frac{1-A}{0.01}\,
      \sqrt{\frac{R\sub{out}}{10^{11}{\rm\,cm}}\,
            \frac{1.4\,M_\odot}{M\sub{NS}}},
\]
where $\eta$ is the ratio of the $(R,z)$ and $(R,\phi)$ viscosities
(see \cite{prin:96}), $\epsilon\equiv{}L\sub{X}/\dot{M}c^2$ the
accretion efficiency, $A$ the disk's \Xray\ albedo, $M\sub{NS}$ the
neutron-star mass, and $R\sub{out}$ the outer disk radius.  WP97 find
that for $F_*\simlt0.1$, the disks are stable; for
$0.1\simlt{}F_*\simlt0.15$, the outer parts become warped and precess
retrogradely, but the inner parts remain flat; for
$0.15\simlt{}F_*\simlt0.2$, the inner parts become warped as well,
precessing progradely; and for $F_*\simgt0.2$, the disks start to
behave somewhat chaotically, precessing less steadily, and often
having inner parts tilted by more than 90 degrees.

The numerical value of $F_*$ in a real system is quite uncertain,
mostly because the disk albedo is poorly known (see WP97), but for
typical \Xray\ binary parameters, one will be close to the unstable
regime.  We will {\em{}assume} that $F_*$ varies, and associate the
switch between spin-up and spin-down state with a switch between a
prograde and retrograde inner disk (for prograde pulsar rotation).
What could drive the variations is unclear, as $F_*$ does not depend
directly on quantities such as $\dot{M}$ which are usually thought to
vary.

The one parameter which might lead to a switch between two states is
the disk albedo.  As discussed by WP97, the albedo likely depends on
the angle of incidence, with $(1-A)\simlt0.1$ for grazing and
$(1-A)\simeq0.4$ for normal incidence.  One could imagine a situation
in which a disk is stable, until a random excursion changes the tilt
such that $F_*$ is raised above the critical value, and the disk
starts to become warped.  This would increase $(1-A)$ further, leading
to further growth, etc.  It remains problematic, however, how one
would reset the system.

\begin{figure}[t]
\epsfxsize0.9\columnwidth\epsfbox{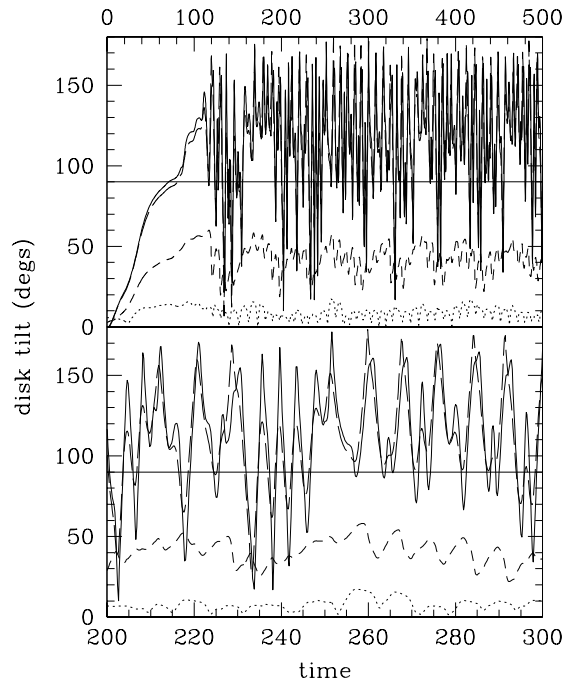}
\caption[]{Disk tilt as a function of time at various radii.  The top
panel shows the full simulation, and the bottom panel a part of it on
an expanded scale.  For the simulation, we used $F_*=0.2$, and matter
enters at two thirds the outer radius (see \cite{wijep:97}).  In units
of the outer radius, the radii shown are at 0.0083, 0.083, 0.47, and
0.87 (solid, long-dash, short-dash, dotted, respectively).  For
\fouru, \CenX3, \oao, and \GX, the outer radii are approximately 0.18,
2, 5, and $50\times10^{11}\un{cm}{}$, respectively.  For
$\dot{M}=10^{-9}\,M_\odot\un{yr}{-1}$, one unit of time equals 0.3,
5, 17, and 300\,d, respectively.\label{fig:tilt}}
\end{figure}

\subsection{Inner-disk tilt}

An example of the time evolution of the disk tilt angle for $F_*=0.2$
is given in Fig.~\ref{fig:tilt}.  It shows that only the inner disk
becomes tilted by more than 90 degrees (i.e., retrograde).  Reversing
the sense of rotation in the inner disk would lead to torque reversals
similar to those observed, just as reversing the whole disk would in
the picture of Nelson et al.\ (\ctyr{nels&a:97}).  A retrograde inner
disk could also lead to the anti-correlation between torque and
luminosity observed in the spin-down state of \GX.

In the model in Fig.~\ref{fig:tilt}, as in most of the simulations we
tried, the average inner disk tilt does not reach 180\arcdeg.  In
contrast, for most values of $F_*$ leading to prograde inner disks,
the tilts are close to 0\arcdeg.  If generally true, then for a
roughly spherical magnetosphere the spin-change rate in the spin-down
state should be systematically lower than that in the spin-up
state\footnote{We note that changes in the orientation of the
neutron-star rotation axis due to the non-parallel component of the
inner-disk angular momentum are likely small.  The effects are reduced
by precession for \GX\ and \fouru, and by randomness in inner disk
orientation for \CenX3 and \oao.}.  For all four sources, the
spin-down rate is indeed smaller, by a factor 1.2 to~4.


Variations around the mean of about 30\arcdeg\ (rms) in the inner disk
tilt angle occur with a characteristic period of $\sim\!4$ time units.
This corresponds to about $10^5\,$s in \fouru, $10^6\,$s in \CenX3,
$10^7\,$s in \oao, and $10^8\,$s in \GX.  (Note, however, that
$F_*=0.2$ is not necessarily appropriate for all systems.)  These
variations result in torque fluctuations, which should contribute to
the torque noise observed in \Xray\ pulsars\footnote{The two
contributions usually considered are $\dot{M}$ variations and internal
torques; see Lamb, Pines, \& Shaham (\ctyr{lambps:78}).}.  For \CenX3
and \oao, this may not be very relevant, as the characteristic period
is comparable to the duration of a spin-up or spin-down state.
However, for \fouru\ and \GX, it is much shorter than the state
duration, and we can compare the noise induced by the fluctuations
with the observations.

For \GX, the tilt variations correspond to a torque noise power of
$\sim\!10^{-16}\un{Hz}{2}\un{s}{-2}\un{Hz}{-1}$ at $10^{-8}\un{Hz}{}$.
This is similar to what is observed (B97).  For \fouru, we need to
integrate the tilt variations over $10^7\,$s, the shortest timescale
on which the torque noise power has been measured.  We find
$\sim\!10^{-21}\un{Hz}{2}\un{s}{-2}\un{Hz}{-1}$, somewhat larger than
the observed $\sim\!10^{-22}\un{Hz}{2}\un{s}{-2}\un{Hz}{-1}$ at
$10^{-7}\un{Hz}{}$ (\cite{chak&a:97a}), but perhaps not inconsistent
given the uncertainties in what a real disk would do.

\subsection{Observing through the disk}

An interesting consequence of associating spin-down with a
flipped-over state is that one should be viewing the neutron star
through the disk at least part of the time.  We estimate the surface
density using a steady, unwarped Shakura-Sunyaev disk.  For all four
systems, $\dot{M}\simeq10^{-9}\,M_\odot\un{yr}{-1}$.  The typical
radius where the line of sight crosses the disk is where the disk tilt
is $90\arcdeg$, about $0.25\,R\sub{out}$ in our simulations.  In
physical units, this ranges between $\sim\!10^{9.5}$ and
$\sim\!10^{12}\,$cm (in \fouru\ and \GX, respectively).  With
$M\sub{NS}=1.4\,M_\odot$ and viscosity parameter $\alpha$, we find a
surface density of $50\alpha^{-4/5}$ and
$0.7\alpha^{-4/5}\un{g}{}\un{cm}{-2}$, respectively (Eq.~5.45 of
\cite{frankr:92}).  This corresponds to column densities $N\sub{H}$ of
about $10^{25}\alpha^{-4/5}$ and $10^{23}\alpha^{-4/5}\un{cm}{-2}$,
respectively.  The disk central temperature due to viscous heating is
$10^{4.5}\alpha^{-1/5}$ and $10^{3}\alpha^{-1/5}\,$K, respectively,
and the ionization parameter $\xi\equiv{}L\sub{X}/nR^2$ a few
$\alpha^{7/10}$ (cgs) for both cases.  Hence, little ionization is
expected, and the disk would be opaque to low-energy X-rays for the
dense case.

In a twisted disk, the extra viscosity associated with the
misalignment between neighboring annuli will lower the column density
somewhat.  Compared to a similar disk with no irradiation, the column
density in the warped disk shown in Figure~\ref{fig:tilt} is lower by
a factor~5 in the outer parts of the disk, and by a factor~2.5 in the
inner parts.

Good data on possible spectral changes associated with the torque
reversals exist only for \fouru\ (Sect.~\ref{sec:obs}).  The observed
decrease and hardening of the 1--20\,keV flux in that source may
reflect a line of sight through the disk.  The spectrum changed mostly
below $\sim\!10\un{keV}{}$, suggesting that absorption might play a
role.  Vaughan \& Kitamoto (\ctyr{vaugk:97}), however, found that the
spectrum could not be fit well by simply increasing the column density
$N\sub{H}$; for an $N\sub{H}$ sufficient to explain the reduction at
close to 10\un{keV}{}, one would expect no soft X-rays whatsoever.

Increasing $N\sub{H}$ may not be quite appropriate, as (i) matter in
the disk is likely neither cold nor homogeneous; (ii) scattering might
be important, as the disk subtends a large solid angle; and (iii) the
obscuration will vary will time, due to precession of the inner disk
-- for suitable inclinations, the source may even be unobscured part
of the time.  Of course, given freedom to change all these, one likely
can fit almost anything.  Vaughan \& Kitamoto could fit the \fouru\
spectrum assuming partial covering, where 86\% of the source is
obscured with $N\sub{H}=10^{23.8}\un{cm}{-2}$ and the remainder
unobscured.  If this were the correct model, it might reflect either
spatial or temporal covering (the spectrum was a long-term average).

\subsection{Problems}

There are several difficulties in applying the warped disk model to
the torque reversal observations in detail.  The primary one concerns
the time scales between the reversals.  In those simulations where a
retrograde inner disk arose, its mean inclination remained relatively
stable (and $>90^\circ$) after its initial growth, with only short
excursions to prograde flow.  This is at odds with the observations,
which show roughly equal times in spin-up and spin-down.

Furthermore, each of the four systems considered has an aspect that is
not easily explained in the retrograde disk picture.  In \fouru, there
is QPO evidence for prograde motion during spin-down, likely at small
radius (Sect.~\ref{sec:obs}).  In \GX, the torque reversals observed
by BATSE were clearly accompanied by luminosity changes.  Finally, in
both high-mass X-ray binaries (Cen X-3 and OAO 1657$-415$), the lack
of a clear correlation between spin-change rate and luminosity (in
both spin-up and spin-down states) seems to confound all the existing
models.

A number of potentially important effects have been neglected in the
WP97 simulations.  These include: the dependence of the albedo on
angle of incidence (and the resulting variations of $F_*$), the
increased disk scale height and thus reduced density due to
irradiation (resulting in larger $\xi$ and possibly different disk
opacity), and the varying ``splash point'' where the accretion stream
-- which is in the orbital plane -- hits the precessing disk
(resulting, e.g., in inward advection in addition to viscosity).
Accounting for these properly may considerably change the detailed
predictions for warped disks. However, the existence of warps and
transitions to retrograde flow under certain conditions seem a general
feature.

\section{Conclusions}\label{sec:conc}

It has been previously suggested that retrograde disk flows would be
an elegant explanation for torque reversals in disk-fed X-ray pulsars.
It has been an open question, however, how such a flow would be
created when mass is transferred by Roche-lobe overflow and thus enters
the accretion disk in a prograde orbit.  We have pointed out that
strong disk warping may produce a retrograde flow close to an
accreting neutron star from an initially prograde flow in the outer
disk.

While this simple picture allows us to understand some of the
characteristics of the observations, it does not match all of them.
This could be because the models currently neglects many potentially
important effects.  Our main point here, however, is that quasi-stable
retrograde disk flows are physically plausible, and that they may thus
be (part of) what causes the spin-down states in persistent disk-fed
X-ray pulsars.

Despite the uncertainties in the models, it remains true generally
that if the spin-down state is due to a flipped-over inner disk, it
should be associated with enhanced absorption.  One would expect
relatively strong absorption edges, especially of iron, which could be
looked for with high-resolution spectroscopy.  The absorption should
vary with time as the inner disk precesses (although the precession is
not necessarily very coherent). With more disk surface exposed to the
central X-ray source, we also would expect increased scattering and
stronger fluorescence lines.

\acknowledgments We thank Mark Finger and Dimitrios Psaltis for useful
discussions.  DC acknowledges support from a NASA Compton GRO
fellowship; RAMJW from a Royal Society URF grant.

\end{document}